\documentclass[prl,amsmath,amssymb,twocolumn]{revtex4}

\usepackage{graphicx}
\usepackage{color}
\usepackage{amsmath}
\usepackage{amssymb}

\usepackage{bm}% bold math
\usepackage{bbm}

\newcommand{\CH}{\mathbb{C}\otimes\mathbb{H}}

\newcommand{\CLsix}{\mathbb{C}l(6)}
\newcommand{\CLseven}{\mathbb{C}l(7)}

\begin{document}

\title{Generations:  Three Prints, in Colour}

\author{Cohl Furey}
\affiliation{$ $ \\ Perimeter Institute for Theoretical Physics, Waterloo, Ontario N2L 2Y5, Canada,\\ University of Waterloo, Ontario N2L 3G1, Canada}
\pacs{12.38.-t, 12.10.Dm, 02.10.Hh, 11.15.-q}

\begin{abstract}

\noindent  We point out a somewhat mysterious appearance of $SU_c(3)$ representations, which exhibit the behaviour of three full generations of standard model particles.  These representations are found in the Clifford algebra $\CLsix$, arising from the complex octonions.  In this paper, we explain how this 64-complex-dimensional space comes about.  With the algebra in place, we then identify generators of $SU(3)$ within it.  These $SU(3)$ generators then act to partition the remaining part of the 64-dimensional Clifford algebra into six triplets, six singlets, and their antiparticles.  That is, the algebra mirrors the chromodynamic structure of exactly three generations of the standard model's fermions.  Passing from particle to antiparticle, or \it vice versa, \rm requires nothing more than effecting the complex conjugate, $*$: $i \mapsto -i$. The entire result is achieved using only the eight-dimensional complex octonions as a single ingredient.
\end{abstract}

\maketitle

\maketitle
\noindent \bf Motivation. \rm  There is more than one way to unify a theory.  In the case of GUTs, unification occurs when the gauge groups of the standard model are engulfed by a single, larger, group.  In other words, \it gauge bosons are unified with gauge bosons. \rm  In the process of pooling these groups together, GUTs manage to merge matter degrees of freedom as well;  \it fermions are unified with fermions. \rm

Classic examples include the packaging of $SU(3)$, $SU(2)$ and $U(1)$ into the 24-dimensional group $SU(5)$.  The numerous representations within a generation are then fused into just two of $SU(5)$'s representations:  the $5^*$ and the $10$.  Alternately, the 45-dimensional ``$SO(10)$" model consolidates the standard model gauge groups, and casts a generation of representations into a single 16-dimensional spinor.

Still, other forms of unified theories can unite other objects.  The early work of G\"{u}naydin and G\"{u}rsey,~\citep{GGquarks}, \citep{GGstats} shows a representation of the Lie algebra $\mathcal{L}G_2$ in terms of sequences of octonions acting on octonions.  Here, \it gauge bosons are unified with fermions. \rm  Extending the work of \citep{GGquarks} was Dixon, \citep{dixon}, who then suggested writing all fermionic degrees of freedom in terms of the tensor product of the division algebras.   \it Local spacetime degrees of freedom are unified with internal degrees of freedom. \rm

Despite the wide range of proposals to simplify the standard model, most schemes tend to share the same impedances.  Few models naturally offer more than a single generation of particles, and few are able to evade proton decay without repercussion.

The purpose of  this article is \it not \rm to offer a completed unified gauge theory, or even a completed description of QCD.  Instead, we expose a gateway from which such a theory might be found.

We come forward with some early blueprints, hinting at  an unusually efficient chromodynamic model.  The bosons here would be drafted from the same algebra as the fermions that they act on.  Better still, this algebra readily supports multiple generations, despite being built from nothing more than the complex octonions: an eight-complex-dimensional algebra.  Paradoxically, it is in fact the \it non-associativity \rm of the octonions that enables a larger \it associative \rm algebra to arise, as peculiar as this may initially sound.

This discovery is expected to strengthen several lines of research.  It may prompt investigators to reinvest in early theories, \citep{GGquarks}-\citep{dixon_fam}, which are based on the idea of division algebras acting on themselves.  It may provide important clues for those working on novel constructions of particle physics~\citep{conlott}-\citep{e6md}.   It also opens up a full arena for study to $G_2$ gauge theory enthusiasts, \citep{rastogi} - \citep{cas}.  Furthermore, this finding releases Unified Theory of Ideals,~\citep{UTI}, from the confines of a single generation, and finally grants anti-particles a space all to their own, which was not a luxury of the original algebra.

\noindent \bf Prerequisite:  $\mathbb{C}\otimes\mathbb{O}$. \rm  For those unfamiliar with the complex octonions, $\mathbb{C} \otimes \mathbb{O}$,  we provide a brief introduction.  It should be noted that all tensor products will be assumed to be over $\mathbb{R}$ in this text, unless otherwise stated.

\begin{figure}[h!]
\includegraphics[width=6cm]{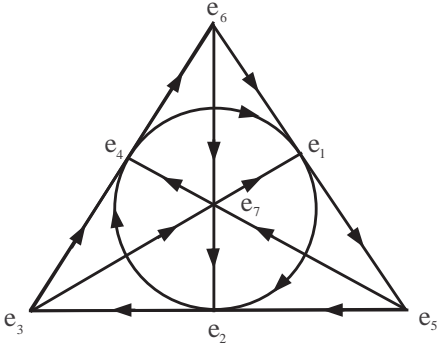}
\caption{\label{fano}
Octonionic multiplication rules}
\end{figure}

The generic element of $\mathbb{C} \otimes \mathbb{O}$ is written $ \sum_{n=0}^7 A_n e_n $, with the $A_n \in \mathbb{C}$.  The $e_n$ are octonionic imaginary units $\left(e_n^2=-1\right)$, apart from $e_0=1$, which multiply according to Figure~\ref{fano}.  The complex imaginary unit $i$ commutes with the octonionic $e_n$.

Any three imaginary units on a directed line segment in Figure~\ref{fano} act as if they were a quaternionic triple (closely related to Pauli matrices).  For example, $e_6e_1 =-e_1e_6= e_5,$ $e_1e_5=-e_5e_1=e_6,$ $e_5e_6=-e_6e_5=e_1,$ $e_4e_1=-e_1e_4=e_2$, etc.  Octonionic multiplication embraces various symmetries, such as index doubling symmetry:  $e_ie_j=e_k \Rightarrow e_{2i}e_{2j}=e_{2k}$, which can be seen by rotating Figure~\ref{fano} by 120 degrees.  For a more thorough introduction to $\mathbb{O}$ see~\citep{baez}, \citep{conway}, \citep{okubo}.

\newpage

\noindent \bf Octonionic chains. \rm  It is plain to see that left-multiplying one complex octonion, $m$, onto another, $f$, provides a map from $f\in\mathbb{C}\otimes\mathbb{O}$ to $f'\equiv mf \in\mathbb{C}\otimes\mathbb{O}$.  Subsequently left-multiplying by another complex octonion, $n$, provides another map: $f\mapsto  f'' \equiv n(mf)$.  We will call this map $\overleftarrow{nm}$, where the arrow is in place so as to indicate the order in which multiplication occurs.  We may extend the chain further by left-multiplying by $p\in\mathbb{C}\otimes\mathbb{O}$, giving $\overleftarrow{pnm}:f\mapsto p(n(mf))$, and so on.

In an associative algebra, the exercise of building up chains in order to make new maps would be futile.  That is, for $m_1, m_2, f$ in an associative algebra, $m_2(m_1f)$ can always be summarized as $m'f$, where $m'\equiv m_2m_1 \in \mathbb{C}\otimes\mathbb{O}$.  However, as the complex octonions form a non-associative algebra, building chains does in fact lead to new maps.  For example, $\overleftarrow{e_{34}}\left(e_6+ie_2 \right) =  e_3 \left( e_4 \left( e_6+ie_2  \right) \right) =  -1 +ie_7 $.  This is not the same as $\left( e_3  e_4 \right) \left( e_6+ie_2  \right)= \left( e_6 \right) \left(e_6+ie_2  \right)=  -1 -ie_7 $, and in fact there exists no $a \in \mathbb{C}\otimes\mathbb{O}$ such that $\overleftarrow{e_{34}}\left( e_6+ie_2 \right) = a \left( e_6+ie_2 \right)$.

Addition and multiplication are easy to define on this set of maps; we will refer to the resulting algebra as the \it octonionic chain algebra, \rm or $\mathbb{C}\otimes\overleftarrow{\mathbb{O}}$.  Addition of two maps $N =\overleftarrow{\cdots n_3 n_2 n_1}$ and $P=\overleftarrow{\cdots p_3 p_2 p_1}$ $\in \mathbb{C}\otimes\overleftarrow{\mathbb{O}}$ on $f$ is given by $\left[N+P\right]f = Nf+Pf,$ where the $n_i$ and $p_j$ $\in \mathbb{C}\otimes\mathbb{O}$.  Multiplication, $\circ$, is given simply by the composition of maps,
\begin{equation}\left[P\circ N\right]f =P(N(f))= \overleftarrow{\cdots p_3 p_2 p_1\cdots n_3 n_2 n_1}f.
\end{equation}
As the composition of maps is always associative, $\mathbb{C}\otimes\overleftarrow{\mathbb{O}}$ is an associative algebra. Unconvinced readers are encouraged to check explicitly that $ \left[\left[A \circ B\right] \circ C \right]f = \left[A \circ \left[B \circ C \right]\right]f$ $\forall$ $A,B,C\in \mathbb{C}\otimes\overleftarrow{\mathbb{O}}$ and $\forall$ $f\in\mathbb{C}\otimes\mathbb{O}$.

Looking more closely at these maps, we notice quickly that
\begin{equation}\label{flip}
\overleftarrow{\cdots e_ae_b\cdots} f = -\hspace{1.5mm}\overleftarrow{\cdots e_be_a\cdots} f \hspace{0.5cm} \forall f \in \mathbb{C}\otimes\mathbb{O},
\end{equation}

\noindent for $a,b = 1,2,\dots 7$, when $a\neq b$.  Furthermore,

\begin{equation}\label{snip}
\overleftarrow{\cdots e_ie_je_je_k\cdots} f = -\hspace{1.5mm}\overleftarrow{\cdots e_ie_k\cdots} f \hspace{0.5cm} \forall f \in \mathbb{C}\otimes\mathbb{O},
\end{equation}
\noindent when $i, k = 0, 1, 2, \dots 7$ and $j=1,2, \dots 7$.  As such, one might suspect an incarnation of the Clifford algebra $\CLseven$, where $\{i\overleftarrow{e_1}, i\overleftarrow{e_2}, \dots i\overleftarrow{e_7}\}$, acting on $f$, form the generating set of vectors.

It turns out, though, that this is not exactly the case.  The chains contain an additional symmetry, which identifies each element of the would-be $\CLseven$ with some other element, thereby cutting $\CLseven$ in half.  For example, $\overleftarrow{e_1e_2e_3}f =-\overleftarrow{e_4e_5e_6e_7}f,$ $\overleftarrow{e_5e_7}f =-\overleftarrow{e_1e_2e_3e_4e_6}f,$ $\overleftarrow{e_7}f =\overleftarrow{e_1e_2e_3e_4e_5e_6}f$, etc.  These 64 equations are readily found by making use of equations (\ref{flip}) and (\ref{snip}), and also the following form of the identity: $\overleftarrow{e_0}f=-\overleftarrow{e_1e_2e_3e_4e_5e_6e_7}f$.  We then see that any element of $\mathbb{C}\otimes\overleftarrow{\mathbb{O}}$ may be represented as a complex linear combination of chains, of no more than three $e_j$s in length.

\begin{figure}[h!]
%\begin{center}
\includegraphics[width=8cm]{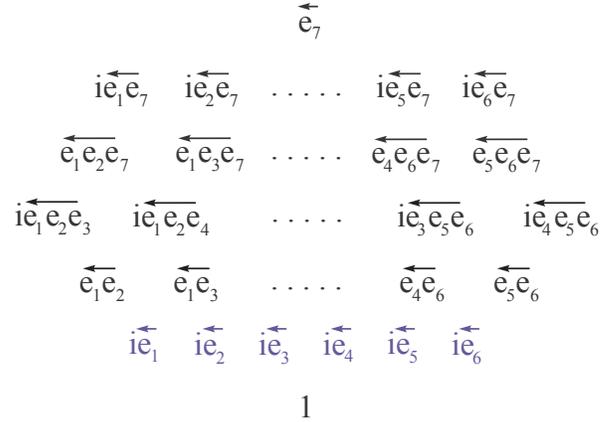}
\caption{\label{cliff6}
The 64-complex-dimensional octonionic chain algebra gives a representation of $\CLsix$.  The octonionic chain algebra is a space of maps acting through left multiplication onto any element $f\in\mathbb{C}\otimes\mathbb{O}$.}
%\end{center}
\end{figure}

The reader is encouraged to check that $\mathbb{C}\otimes\overleftarrow{\mathbb{O}}$ forms the 64-complex-dimensional Clifford algebra $\CLsix$, generated by the set $\{i\overleftarrow{e_1}, i\overleftarrow{e_2}, \dots i\overleftarrow{e_6}\}$, acting on $f$.  Figure~(\ref{cliff6}) depicts the octonionic chain algebra, organized so as to demonstrate its $\CLsix$ structure.  Starting from the bottom, we have the zero-vector, $1$ acting on $f$, the 1-vectors, $\{i\overleftarrow{e_1}, i\overleftarrow{e_2}, \dots i\overleftarrow{e_6}\}$ acting on $f$, the 2-vectors, $\{\overleftarrow{e_1e_2},\dots  \overleftarrow{e_5e_6}\}$ acting on $f$, and so on.  Note that we make regular use the identity $\overleftarrow{e_7}f=\overleftarrow{e_1e_2e_3e_4e_5e_6}f$ so as to avoid writing long chains of multivectors involving only the generators $i\overleftarrow{e_1}, \hspace{.7mm}i\overleftarrow{e_2}, \dots i\overleftarrow{e_6}$.  For earlier work which makes reference to the octonionic chain algebra, see \citep{dixon} and \citep{Leo}.

\newcommand{\spacer}{0.2cm}
\noindent \bf $SU(3)$'s Lie Algebra. \rm  A generic element, $\Lambda$, of $su(3)$, acting on $f\in\mathbb{C}\otimes\mathbb{O}$ may be expressed as
\begin{equation}\begin{array}{ll}\label{liealg}
\Lambda f=&\sum_{k=1}^8\lambda_k\Lambda_kf\equiv\big[\lambda_1\frac{i}{2}[\overleftarrow{e_{15}}-\overleftarrow{e_{34}}]+\lambda_2\frac{i}{2}[-\overleftarrow{e_{14}}
-\overleftarrow{e_{35}}]\vspace{\spacer}\\
&+\lambda_3\frac{i}{2}[-\overleftarrow{e_{13}}+\overleftarrow{e_{45}}]
+\lambda_4\frac{i}{2}[\overleftarrow{e_{25}}+\overleftarrow{e_{46}}]\vspace{\spacer}
\\ &+\lambda_5\frac{i}{2}[-\overleftarrow{e_{24}}+\overleftarrow{e_{56}}]+\lambda_6\frac{i}{2}[\overleftarrow{e_{16}}+\overleftarrow{e_{23}}]
\vspace{\spacer}
\\ &+\lambda_7\frac{i}{2}[\overleftarrow{e_{12}}+\overleftarrow{e_{36}}]
+\lambda_8\frac{i}{2\sqrt{3}}[\overleftarrow{e_{13}}+\overleftarrow{e_{45}}-2\overleftarrow{e_{26}}]\big]f,\vspace{\spacer}
\end{array}\end{equation}
\noindent  where $\overleftarrow{e_{ab}}$ is shorthand for $\overleftarrow{e_{a}e_b}$.  Note that we will take the $\lambda_k$ to be in $\mathbb{C}$.  As one would expect, the commutation relations take the form
\begin{equation}\label{comm}
\Bigg[ \frac{\Lambda_a}{2} \hspace{0.5mm} ,\hspace{0.5mm} \frac{\Lambda_b}{2} \Bigg] f \equiv \left[ \frac{\Lambda_a}{2} \circ \frac{\Lambda_b}{2} -\frac{\Lambda_b}{2} \circ \frac{\Lambda_a}{2} \right]f= ic_{abc} \frac{\Lambda_c}{2} f,
\end{equation}
\noindent $\forall f \in \mathbb{C}\otimes\mathbb{O}$, with the usual $SU(3)$ structure constants $c_{abc}$.  The above representation of $su(3)$ can be found in~\citep{GGquarks}, introduced as a subalgebra of $\mathcal{L}G_2$ acting on the octonions.

Clearly, $\Lambda$, as expressed above, constitutes an element of the chain algebra, $\mathbb{C}\otimes\overleftarrow{\mathbb{O}}$.  In earlier references, ~\citep{GGquarks}, \citep{GGstats}, \citep{dixon},  this $\Lambda$ is shown to act on quark and lepton representations in the eight-dimensional $\mathbb{C}\otimes\mathbb{O}$, or multiple copies thereof.  In contrast, here we introduce $\Lambda$ acting on quark and lepton representations of the 64-dimensional $\mathbb{C}\otimes\overleftarrow{\mathbb{O}}$.

Taking a hint from~\citep{UTI}, let us now introduce a related representation of $su(3)$, which will draw out structure in $\mathbb{C}\otimes\overleftarrow{\mathbb{O}}$, familiar from the behaviour of quarks and leptons.

Consider a resolution of the identity in $\mathbb{C}\otimes\overleftarrow{\mathbb{O}}$
\begin{equation} 1f = \left[\nu + \nu^*\right]f,
\end{equation}
\noindent where $\nu \equiv \frac{1}{2}(1+i\overleftarrow{e_7})$.  Both $\nu$ and $\nu^*$ act as projectors, whereby $\nu\circ\nu = \nu$, $\hspace{2mm}\nu^*\circ\nu^* = \nu^*$, and $\hspace{1mm}\nu\circ\nu^*=\nu^*\circ\nu=0$.  For those familiar with ideals, it is straightforward to see that objects of the form $a\circ\nu$ form a non-trivial ideal under left multiplication, where $a\in\mathbb{C}\otimes\overleftarrow{\mathbb{O}}$.  The same can be said for objects of the form $a\circ\nu^*$.

In order to be rid of some redundant notation, we will now abandon the use of the symbol $\circ$ for multiplication, replacing it instead with concatenation.  Also, it is to be assumed from this point forward that equations between elements of $\mathbb{C}\otimes\overleftarrow{\mathbb{O}}$ hold over all $f\in\mathbb{C}\otimes\mathbb{O}$, even though $f$ will now be absent from the equations.

As \begin{equation} \big[\Lambda_n \hspace{0.5mm} ,\hspace{0.5mm} \nu \big] = 0 \hspace{1cm} \forall n=1\dots8,
\end{equation}
\noindent equation~(\ref{comm}) then leads to
\begin{equation}\label{commnu}
\Bigg[ \frac{\Lambda_a}{2}\nu \hspace{0.5mm} ,\hspace{0.5mm} \frac{\Lambda_b}{2}\nu \Bigg] = ic_{abc} \frac{\Lambda_c}{2}\nu.
\end{equation}
\noindent That is, the eight $\frac{1}{2}\Lambda_n\nu$ form a representation of $su(3)$.  Taking the complex conjugate of~(\ref{commnu}) gives
\begin{equation}\label{commnustar}
\Bigg[ -\frac{\Lambda_a^*}{2}\nu^* \hspace{0.5mm} ,\hspace{0.5mm} -\frac{\Lambda_b^*}{2}\nu^* \Bigg] = ic_{abc} \left[-\frac{\Lambda_c^*}{2}\nu^*\right],
\end{equation}
\noindent so that the $-\frac{1}{2}\Lambda_n^*\nu^*$ give a further representation.

\noindent \bf Families of Colour. \rm  Knowing that the $\Lambda_n\nu$ behave as an eight dimensional representation under the action of $\left[\Lambda_m\nu \hspace{0.5mm} , \hspace{1.5mm}\cdot\hspace{1.5mm}\right]$, one might wonder how objects of the more general form $a\nu$ behave under $\left[\Lambda_m\nu \hspace{0.5mm} ,\hspace{1.5mm}\cdot\hspace{1.5mm}\right]$.

Obeying $\left[\Lambda_m\nu\hspace{0.5mm} ,\hspace{0.5mm} \ell_j\nu\right]=0$ $\forall m=1\dots8$, we find six $SU(3)$ singlets, whose basis vectors are given by
\begin{equation}\begin{array}{l}
\ell_{a}\equiv\nu, \hspace{.8cm}  \ell_{b}\equiv\left(e_{13}+e_{26}+e_{45}\right)\nu, \vspace{\spacer}\\
\ell_{c}\equiv \left(-ie_{124}-e_{125}+e_{146}-ie_{156}\right)\nu, \vspace{\spacer} \\
\ell_{d}\equiv \left(-ie_{1}-e_{3}+e_{126}+e_{145}\right)\nu, \vspace{\spacer} \\
\ell_{e}\equiv \left(ie_{2}+e_{6}+e_{123}+ie_{136}\right)\nu, \vspace{\spacer} \\
\ell_{f}\equiv \left(ie_{4}+e_{5}-e_{134}+ie_{135}\right)\nu, \vspace{\spacer} \\
\end{array}\end{equation}
\noindent where the left-pointing arrows were dropped throughout for notational simplicity, and right-to-left multiplication is still meant to occur. The set of basis vectors
\begin{equation}\begin{array}{l}\label{triplet}
q^R_{1}\equiv\left(-ie_{12}-e_{16}+e_{23}+ie_{36}\right)\nu   \vspace{\spacer} \\
q^G_{1}\equiv\left(-ie_{24}-e_{25}+e_{46}-ie_{56}\right)\nu \vspace{\spacer} \\
q^B_{1}\equiv\left(ie_{14}+e_{15}+e_{34}-ie_{35}\right)\nu \vspace{\spacer} \\
\end{array}\end{equation}
\noindent acts as a triplet under commutation with the $\Lambda_m\nu$.  Next, we find five anti-triplets given by
\begin{equation}\begin{array}{l}
\bar{q}^R_{2}\equiv\left(ie_{12}-e_{16}+e_{23}-ie_{36}\right)\nu   \vspace{\spacer} \\
\bar{q}^G_{2}\equiv\left(ie_{24}-e_{25}+e_{46}+ie_{56}\right)\nu \vspace{\spacer} \\
\bar{q}^B_{2}\equiv\left(-ie_{14}+e_{15}+e_{34}+ie_{35}\right)\nu, \vspace{\spacer} \\
\end{array}\end{equation}
\begin{equation}\begin{array}{l}
\bar{q}^R_{3}\equiv\left(ie_{4}+e_{5}+e_{134}-ie_{135}\right)\nu   \vspace{\spacer} \\
\bar{q}^G_{3}\equiv\left(ie_{1}+e_{3}+e_{126}+e_{145}\right)\nu   \vspace{\spacer} \\
\bar{q}^B_{3}\equiv\left(ie_{2}+e_{6}-e_{123}-ie_{136}\right)\nu,   \vspace{\spacer} \\
\end{array}\end{equation}
\begin{equation}\begin{array}{l}
\bar{q}^R_{4}\equiv\left(ie_{1}-e_{3}+e_{126}-e_{145}\right)\nu   \vspace{\spacer} \\
\bar{q}^G_{4}\equiv\left(-ie_{4}+e_{5}+e_{134}+ie_{135}\right)\nu   \vspace{\spacer} \\
\bar{q}^B_{4}\equiv\left(ie_{124}-e_{125}-e_{146}-ie_{156}\right)\nu,   \vspace{\spacer} \\
\end{array}\end{equation}
\begin{equation}\begin{array}{l}
\bar{q}^R_{5}\equiv\left(-ie_{2}+e_{6}+e_{123}-ie_{136}\right)\nu   \vspace{\spacer} \\
\bar{q}^G_{5}\equiv\left(ie_{124}-e_{125}+e_{146}+ie_{156}\right)\nu   \vspace{\spacer} \\
\bar{q}^B_{5}\equiv\left(ie_{4}-e_{5}+e_{134}+ie_{135}\right)\nu,   \vspace{\spacer} \\
\end{array}\end{equation}
\begin{equation}\begin{array}{l}
\bar{q}^R_{6}\equiv\left(ie_{124}+e_{125}+e_{146}-ie_{156}\right)\nu   \vspace{\spacer} \\
\bar{q}^G_{6}\equiv\left(ie_{2}-e_{6}+e_{123}-ie_{136}\right)\nu   \vspace{\spacer} \\
\bar{q}^B_{6}\equiv\left(-ie_{1}+e_{3}+e_{126}-e_{145}\right)\nu.   \vspace{\spacer} \\
\end{array}\end{equation}

Taking the complex conjugate, $*$: $i\mapsto-i$, of these 32 basis vectors gives 32 new linearly independent basis vectors. Under commutation with $-\Lambda_m^* \nu^*$,
\begin{equation}\begin{array}{l}
\ell_{a}^*=\nu^*, \hspace{.8cm}  \ell_{b}^*=\left(e_{13}+e_{26}+e_{45}\right)\nu^*, \vspace{\spacer}\\
\ell_{c}^*= \left(ie_{124}-e_{125}+e_{146}+ie_{156}\right)\nu^*, \vspace{\spacer} \\
\ell_{d}^*= \left(ie_{1}-e_{3}+e_{126}+e_{145}\right)\nu^*, \vspace{\spacer} \\
\ell_{e}^*= \left(-ie_{2}+e_{6}+e_{123}-ie_{136}\right)\nu^*, \vspace{\spacer} \\
\ell_{f}^*= \left(-ie_{4}+e_{5}-e_{134}-ie_{135}\right)\nu^* \vspace{\spacer} \\
\end{array}\end{equation}
\noindent transform as $SU(3)$ singlets,
\begin{equation}\begin{array}{l}
q^{R*}_{1}=\left(ie_{12}-e_{16}+e_{23}-ie_{36}\right)\nu^* \equiv\bar{q}^R_1   \vspace{\spacer} \\
q^{G*}_{1}=\left(ie_{24}-e_{25}+e_{46}+ie_{56}\right)\nu^* \equiv\bar{q}^G_1 \vspace{\spacer} \\
q^{B*}_{1}=\left(-ie_{14}+e_{15}+e_{34}+ie_{35}\right)\nu^* \equiv\bar{q}^B_1 \vspace{\spacer} \\
\end{array}\end{equation}
\noindent behaves as an anti-triplet,
\begin{equation}\begin{array}{l}
\bar{q}^{R^*}_{2}=\left(-ie_{12}-e_{16}+e_{23}+ie_{36}\right)\nu^*\equiv q_2^R   \vspace{\spacer} \\
\bar{q}^{G^*}_{2}=\left(-ie_{24}-e_{25}+e_{46}-ie_{56}\right)\nu^*\equiv q_2^G \vspace{\spacer} \\
\bar{q}^{B^*}_{2}=\left(ie_{14}+e_{15}+e_{34}-ie_{35}\right)\nu^*\equiv q_2^B \vspace{\spacer} \\
\end{array}\end{equation}
\noindent behaves as a triplet, and so on.

That is, unlike the standard model, we are able to pass back and forth between particle and anti-particle using \it only \rm the complex conjugate $i\mapsto-i$.   This feature appeared early on in the work of~\citep{GGquarks} for some  internal degrees of freedom, and also in~\citep{UTI} when passing between left- and right-handed Weyl spinors.  In the case of~\citep{UTI}, we showed that the well-known $2\times2$ matrix, $\epsilon$, is made obsolete in our formalism.

\noindent \bf A sample calculation. \rm  We introduce to the reader how calculations are carried out in $\mathbb{C}\otimes\overleftarrow{\mathbb{O}}$  by working through an example.  Let us find the action of the first $SU(3)$ generator of the form $\Lambda \nu$, which we will define as $\Lambda_1\nu\equiv\frac{i}{2}\left(e_{15}-e_{34}\right)\nu$, in accordance with equation~(\ref{liealg}).  Let $\Lambda_1\nu$ act on $q^R_{1}$, as defined in equations~(\ref{triplet}):
\begin{equation}\begin{array}{ll}\label{calc} & \Big[ \Lambda_1\nu \hspace{1mm},\hspace{1mm} q^R_{1} \Big]\vspace{\spacer}   \\
=& \Big[\frac{i}{2}\left(e_{15}-e_{34}\right)\nu\hspace{0.5mm} ,\hspace{1mm}\left(-ie_{12}-e_{16}+e_{23}+ie_{36}\right)\nu \Big]\vspace{\spacer} \\
=&\frac{i}{2}\Big( \left(e_{15}-e_{34}\right)\left(-ie_{12}-e_{16}+e_{23}+ie_{36}\right)\vspace{\spacer} \\
 &- \left(-ie_{12}-e_{16}+e_{23}+ie_{36}\right)\left(e_{15}-e_{34}\right) \Big)\hspace{0.2mm}\nu\vspace{\spacer} \\
%\end{array}\end{equation}
%\begin{equation}\begin{array}{ll}
=&\frac{i}{2}\big(-ie_{1512}-e_{1516}+e_{1523}+ie_{1536}\vspace{\spacer} \\
&\hspace{0.85cm}+ie_{3412}+e_{3416}-e_{3423}-ie_{3436}\vspace{\spacer} \\
&\hspace{0.85cm}+ie_{1215}+e_{1615}-e_{2315}-ie_{3615}\vspace{\spacer} \\
&\hspace{0.85cm}-ie_{1234}-e_{1634}+e_{2334}+ie_{3634}\big)\hspace{0.2mm}\nu\vspace{\spacer} \\
\end{array}\end{equation}
\begin{equation}\begin{array}{ll}
=&\frac{i}{2}\big(-ie_{52}-e_{56}+e_{1235}-ie_{1356}\vspace{\spacer} \\
&\hspace{0.85cm}+ie_{1234}+e_{1346}+e_{42}-ie_{46}\vspace{\spacer} \\
&\hspace{0.85cm}+ie_{25}+e_{65}-e_{1235}+ie_{1356}\vspace{\spacer} \\
&\hspace{0.85cm}-ie_{1234}-e_{1346}-e_{24}+ie_{64}\big)\hspace{0.2mm}\nu\vspace{\spacer} \\
=&i\big(ie_{25}-e_{56}-e_{24}-ie_{46}\big)\hspace{0.2mm}\nu=q^G_{1}.\vspace{\spacer} \\
\end{array}\end{equation}

\noindent This is the result we would expect for the first of the $su(3)$ Gell-Mann matrices, $\Lambda^{GM}_1$, from the standard model, acting to convert a red basis vector, $\underline{R}\equiv (1,0,0)^\top$, into a green basis vector, $\underline{G}\equiv (0,1,0)^\top$.
\begin{equation} \Lambda^{GM}_1\underline{R} = \left(
\begin{array}{ccc}
0 & 1 & 0 \\
1 & 0 & 0 \\
0 & 0 & 0
\end{array} \right)\left(
\begin{array}{c}
1  \\
0  \\
0
\end{array} \right) = \left(
\begin{array}{c}
0  \\
1  \\
0
\end{array} \right) = \underline{G}.
\end{equation}

\noindent \bf A bigger picture.  \rm  Unified Theory of Ideals,~\citep{UTI}, suggested a new model for describing particle representations in terms of the tensor product of the division algebras, $\mathbb{R}\otimes\mathbb{C}\otimes\mathbb{H}\otimes\mathbb{O}=\mathbb{C}\otimes\mathbb{H}\otimes\mathbb{O}$.  The \it Dixon algebra, \rm as it is called, is the tensor product of the reals, $\mathbb{R}$, the complex numbers, $\mathbb{C}$, the quaternions, $\mathbb{H}$, and the octonions, $\mathbb{O}$.  In~\citep{UTI}, Dirac spinors $\Psi_L+\Psi_R$, were represented by the complex quaternions, $\mathbb{C}\otimes\mathbb{H}$.  Internal degrees of freedom for one full generation of particles:  an up-type colour-triplet, down-type colour-triplet, a neutrino and an electron, were identified with the eight-dimensional algebra $\mathbb{C}\otimes\mathbb{O}$.  Fitting $\mathbb{C}\otimes\mathbb{H}$ and $\mathbb{C}\otimes\mathbb{O}$ together via a tensor product over $\mathbb{C}$ gives the Dixon algebra.

In this paper, we are now suggesting to replace this internal space $\mathbb{C}\otimes\mathbb{O}$ with $\mathbb{C}\otimes\overleftarrow{\mathbb{O}}$.  The completed space, $\mathbb{C}\otimes\mathbb{H}\otimes\overleftarrow{\mathbb{O}}=\overleftarrow{\mathbb{C}}\otimes\overleftarrow{\mathbb{H}}\otimes\overleftarrow{\mathbb{O}}$, then assigns a left- and right-handed spinor to each member of the colour triplets, anti-triplets, and singlets of $\mathbb{C}\otimes\overleftarrow{\mathbb{O}}$.  In particular, this is true for the $SU(3)$ singlets, where neutrinos and anti-neutrinos could reside.  That is, as with \citep{dixon} and \citep{UTI}, this model provides for the existence of a right-handed neutrino.  In the case of gauge bosons, $\CH$ was shown in \citep{UTI} to be capable of also describing four-vectors, so that the basic framework is available for their polarizations.

\begin{figure}[h!]
\includegraphics[width=8.5cm]{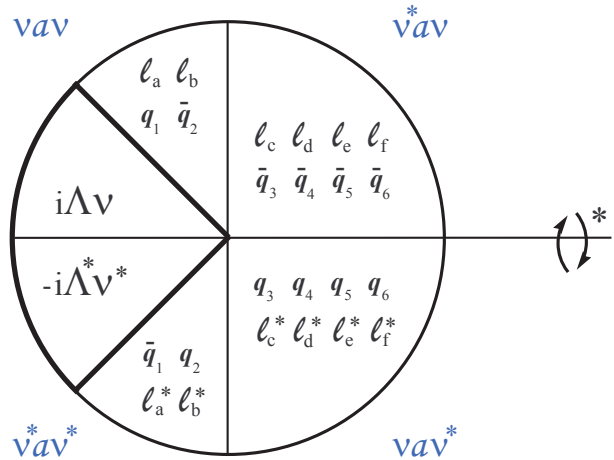}
\caption{\label{64} The 64-dimensional octonionic chain algebra splits into two sets of $SU(3)$ generators of the form $i\Lambda\nu$ and $-i\Lambda^*\nu^*$, six $SU(3)$ singlets $\ell_j$, six triplets $q_k$, and their complex conjugates.  These objects are sectioned off above into four quadrants according to their forms: $\nu a \nu$, $\nu^* a \nu$, $\nu a \nu^*$ and $\nu^* a \nu^*$ for $a$ in the chain algebra. Transforming particles into anti-particles, and \it vice versa, \rm requires only the complex conjugate $*$: $i\mapsto-i$ in our formalism.}
\end{figure}

Finally, we conclude by summarizing the main result of this paper in Figure~(\ref{64}): the breakdown of the 64-dimensional $\mathbb{C}\otimes\overleftarrow{\mathbb{O}}$ into irreducible representations of $SU(3)$.

\noindent \bf Open Questions. \rm  From these results immediately follow a couple of questions:  \it What about the standard model's electro-weak sector? \rm This is a question which should be approached with caution.  That is, experience with division algebras teaches quickly that $SU(2)$ and $U(1)$ symmetries appear in abundance in these spaces.  This makes it easy to be misled into thinking that the correct $SU(2)$ and $U(1)$ symmetries have been identified, when in fact they have not.

\it What is $\nu$? \rm Projectors appearing in a model without full justification for their existence should provoke suspicion.  In this case, the projector $\nu$ came about as a result of the \it complex multiplicative action \rm of~\citep{UTI}.  It is believed that $\nu$ could further be tied in with the formation of a Jordan algebra, an idea which is  currently under investigation.

%However, $\nu=\frac{1}{2}\left(1+i\overleftarrow{e_7}\right)$ was not the only choice of projector available.  It remains to be seen how changes in the choice of $\nu$, and how $\nu$ is applied, will affect the various representations of $SU(3)$.

%Furthermore, a conjugate scheme, $ \left[ \Lambda \nu, \nu a \right]$ and $\left[-\Lambda ^* \nu ^*, \nu ^* a\right]$, also gives the same chromodynamic structure as was found is this paper via $ \left[ \Lambda \nu, a\nu  \right]$ and $\left[-\Lambda ^* \nu ^*, a\nu ^* \right]$.  

\noindent \bf Conclusion. \rm  Using only the eight-dimensional complex octonions, $\mathbb{C}\otimes\mathbb{O}$, we have explained how to build up a 64-complex-dimensional associative algebra.  The $SU(3)$ generators identified within this algebra then break down the remaining space into six singlets, six triplets, and their antiparticles, with no extra particles beyond these.  

These representations are a curious finding.  They effortlessly suggest the existence of exactly three generations, they relate particles to antiparticles by using only the complex conjugate $i\mapsto -i$, and finally, they fill these tall orders while working from but a modest eight-complex-dimensional algebra.

\noindent \bf Acknowledgements. \rm \it To Allan Furey, who saw how to make something from almost nothing.\rm

The author is indebted to S. Farnsworth, G. Fiore,  L. Smolin, and C. Tamarit for their feedback on this work.  This research was supported by the Templeton Foundation, and also in part by Perimeter Institute for Theoretical Physics. Research at Perimeter Institute is supported by the Government of Canada through Industry Canada and by the Province of Ontario through the Ministry of Research and Innovation.

\end{document}